\documentclass{article}

\usepackage[table]{xcolor}
\usepackage[margin=1in]{geometry}
\usepackage{tabularx}
\usepackage{enumitem}
\usepackage{longtable,tabu}
\setlist{nolistsep}
\usepackage{array,booktabs}

\usepackage[utf8]{inputenc} 
\usepackage[T1]{fontenc}    
\usepackage{hyperref}       
\usepackage{url}            
\usepackage{booktabs}       
\usepackage{amsfonts}       
\usepackage{nicefrac}       
\usepackage{microtype}      

\usepackage{multirow}
\usepackage{booktabs}

\usepackage{longtable}
\usepackage{threeparttable}
\usepackage{textcomp}
\usepackage{array}

\usepackage{lipsum}

\newcommand\blfootnote[1]{%
  \begingroup
  \renewcommand\thefootnote{}\footnote{#1}%
  \addtocounter{footnote}{-1}%
  \endgroup
}

\title{Acoustic Correlates of the Voice Qualifiers: A Survey\blfootnote{Preprint. Work in progress.}}

\author{
  Shahan Ali Memon\\
  Language Technologies Institute \\
  Carnegie Mellon University \\
  Pittsburgh, PA 15213 \\
  \texttt{\{samemon\}}@cs.cmu.edu
}
\date{}
\begin{document}

\maketitle

\begin{abstract}
Our voices are as distinctive as our faces and fingerprints. There is a spectrum of non-disjoint traits that make our voices unique and identifiable, such as the fundamental frequency, the intensity, and most interestingly the quality of the speech. Voice quality refers to the characteristic features of an individual's voice.

Previous research has from time-to-time proven the ubiquity of voice quality in making different paralinguistic inferences. These inferences range from identifying personality traits, to health conditions and beyond.

In that context, one could arguably summarize the constituent perceptual elements of voice quality into two categories: primary qualities, and voice qualifiers. Primary qualities, as defined by Poyatos \cite{poyatos1991paralinguistic}, are well-defined traits that are always present in the human voice. These include timbre, loudness, pitch, intonation range, etc.

However, what makes our voices truly unique is a spectrum of complex \emph{voice qualifiers} which differ significantly across different speakers and may in fact define different speaker states and traits.

In loose terms, Dr. Ingo Titze defines these voice qualifiers as proxy features in determination of ones' genetic and anatomical structure, and the learned components. In the 8th Vocal Fold Physiology Conference in April 1994, he proposed a sufficiently comprehensive list of voice qualifiers which we use as basis for our research.

In this manuscript, we first map the paralinguistic voice qualifiers to their acoustic correlates in the light of the previous research and literature. We also determine the openSMILE correlates one could possibly use to measure those correlates. In the second part, we give a set of example paralinguistic inferences that can be made using different acoustic and perceptual voice quality features.

\end{abstract}
\pagebreak
\section{Acoustic Correlates of Voice Quality}
\label{sec:vqtable}
Table~\ref{tab:1} maps each of the voice quality features to its definition, and the acoustic correlates found in the background literature found as useful cues to measure it. We also provide the possible openSMILE\cite{eyben2010opensmile} features that could be used as proxy features in relation to the acoustic features. 
\begin{longtable}{
p{1.5cm}
p{2.0cm}
p{2.0cm}
p{4.0cm}
p{5.0cm}
}
\caption{Acoustic Correlates of Voice Quality}\label{tab:1}\\
 \toprule  
\multicolumn{1}{c}{Voice Quality}&
\multicolumn{1}{c}{Perceptive Definition}&
\multicolumn{1}{c}{Physiology}&
\multicolumn{1}{c}{Correlate(s)}&
\multicolumn{1}{c}{openSMILE Correlate(s)}\\
\midrule
\endfirsthead
\caption[]{Acoustic Correlates of Voice Quality}\\
 \toprule  
\multicolumn{1}{c}{Voice Quality}&
\multicolumn{1}{c}{Perceptive Definition}&
\multicolumn{1}{c}{Physiology}&
\multicolumn{1}{c}{Correlate(s)}&
\multicolumn{1}{c}{openSMILE Correlate(s)}\\ \midrule
\endhead

Aphonic & no sound or whisper & inability to set vocal folds into vibration, caused by lack of appropriate power (air pressure) or a muscular/tissue problem of the folds & 
\begin{itemize}[label={--},noitemsep,leftmargin=*,topsep=0pt,partopsep=0pt]
    \item Pitch Absent
    \item Lower HNR
\end{itemize} & 
\begin{itemize}[label={--},noitemsep,leftmargin=*,topsep=0pt,partopsep=0pt]
    \item F0final\_sma\_amean
    \item logHNR\_sma\_amean
\end{itemize}
\\
\hline
Biphonic & two independent pitches & two sources of sound (e.g., true folds and false folds, or two folds and whistle due to vortex in air) &

\begin{itemize}[label={--},noitemsep,leftmargin=*,topsep=0pt,partopsep=0pt]
    \item F1 within 10\% of pitch
    \item Roughness \cite{keating2015acoustic}
\end{itemize} & 
\begin{itemize}[label={--},noitemsep,leftmargin=*,topsep=0pt,partopsep=0pt]
	\item F0final\_sma\_amean
    \item F1frequency\_sma3nz\_amean
    \item openSMILE correlates for Roughness from ~\ref{tab:1}
\end{itemize}
\\
\hline
Breathy & sound of air is apparent or impression of glottal air leakage and turbulence noise during phonation & noise is caused by turbulence in or near glottis, caused by loose valving of laryngeal muscles (lateral cricoarytenoid, interarytenoid and posterior cricoarytenoid) & 
\begin{itemize}[label={--},noitemsep,leftmargin=*,topsep=0pt,partopsep=0pt]
    \item Pitch Amplitude (PA) - supposed to be lower for voiced sounds since PA represents the degree of voicing \cite{doi:10.1044/jshr.3302.298}
    \item \%Jitter which is the mean jitter in ms divided by the mean period in ms - found to be higher for breathy voices. \cite{doi:10.1044/jshr.3302.298}
    \item High Peak levels of the first harmonic \cite{de1995some}
    \item Slope of log-magnitude spectrum in 2-5 KHz band; the steepness of the spectral slope is strongly related to the glottal closure; the faster the glottis is closed, the more pulse like excitation resulting in flat harmonic source spectrum. Breathy voices are gradual - with insufficient glottal closure, yielding spectral slope more than 12 dB per octave. \cite{de1995some}\cite{childers1991vocal}
    \item HNR in b1 and b2 \cite{de1995some}\cite{wolfe1987prediction}
    \item F0 estimated in the cepstral domain \cite{stranik2014acoustic}
    \item glottal-to-noise excitation ratio
\cite{stranik2014acoustic}
	\item high-to-mid or low-frequency energy
\cite{stranik2014acoustic}
	\item ratio of number of voiced parts to number of all samples in the utterance \cite{stranik2014acoustic}
    \item Energy in 5-8 KHz band $\equiv$ Energy in 2-5 KHz band for breathy utterances; non-breathy utterances exhibit lower energy in 5-8 KHz region. \cite{labuschagne2016perception}\cite{hammarberg1980perceptual}
    \item H1-H2 \cite{hillenbrand1994acoustic}\cite{labuschagne2016perception}\cite{klatt1990analysis}
    \item Higher $\frac{H}{L}$ found for breathy voices \cite{hillenbrand1996acoustic}
  \end{itemize} &
\begin{itemize}[label={--},noitemsep,leftmargin=*,topsep=0pt,partopsep=0pt]
    \item jitterLocal\_sma\_amean
    \item logHNR\_sma\_amean
    \item F0semitoneFrom27.5Hz
    \item F0final\_sma\_amean
    \item logRelF0-H1-H2\_sma3nz\_amean
    \item voicingFinalUnclipped
    \item pcm\_RMSenergy
\end{itemize}\\
  \\

\hline
Covered & muffled or 'darkened' sound  & lips are rounded and protruded or larynx is lowered to lower all formants so a stronger fundamental is obtained &
\begin{itemize}[label={--},noitemsep,leftmargin=*,topsep=0pt,partopsep=0pt]
    \item High Spectral Entropy
    \item Lower F1,F2, and F3
    \item Stronger F0
\end{itemize} &
\begin{itemize}[label={--},noitemsep,leftmargin=*,topsep=0pt,partopsep=0pt]
    \item pcm\_fftMag\_spectralEntropy
    \item F1frequency\_sma3nz\_amean
    \item F2frequency\_sma3nz\_amean
    \item F3frequency\_sma3nz\_amean
    \item F0final\_sma3nz\_amean
\end{itemize}\\
\hline
Creaky & sounding like two hard surfaces rubbing against one another & a complex pattern of vibrations in the vocal folds creates a intricate formation of subharmonics and modulations & 
\begin{itemize}[label={--},noitemsep,leftmargin=*,topsep=0pt,partopsep=0pt]
    \item Lower F0 \cite{garellek2012timing}\cite{gerfen2005production}
    \item irregular F0 \cite{keating2015acoustic}
    \item higher values of SHR (subharmonics-to-harmonic ratio) \cite{garellek2015perception}
    \item relatively weak H1 (due to low airflow through constricted glottis)
    \item low B1 values \cite{garellek2015perception}
    \item relatively strong frequency harmonics (which together result in lower values of various harmonic difference measures (H1-A1, H1-A2, H1-A3, H2-H4, H4-2k, 2K-5k) \cite{keating2015acoustic}
    \item  low H1-H2 \cite{garellek2015perception}
    \item less spectral tilt (low flow through glottis means less energy in H1, So harmonic amplitude difference have lower values hence less spectral tilt) \cite{garellek2015perception}
\end{itemize} &
\begin{itemize}[label={--},noitemsep,leftmargin=*,topsep=0pt,partopsep=0pt]
    \item F0final\_sma\_amean
    \item jitterLocal\_sma\_amean
    \item logRelF0-H1-H2\_sma3nz\_amean
    \item logRelF0-H1-A3\_sma3nz\_amean
    \item pcm\_fftMag\_mfcc\_sma[0]
    \item pcm\_fftMag\_mfcc\_sma[1]
    \item pcm\_fftMag\_mfcc\_sma[2]
    \item pcm\_fftMag\_mfcc\_sma[3]
\end{itemize}\\
\hline
Diplophonic & pitch supplemented with another pitch one octave lower, roughness usually apparent & a period doubling, or F0/2 subharmonic & Non-pressed creaky voice \cite{ishi2006acoustic} & openSMILE correlates for pressed and creaky voice in ~\ref{tab:1}\\
\hline
Flutter & often called bleat because it sounds like a lamb's cry & amplitude changes or frequency modulations in the 8-12Hz range &  
\begin{itemize}[label={--},noitemsep,leftmargin=*,topsep=0pt,partopsep=0pt]
    \item Percent Jitter \cite{ncvs.org}
    \item Percent Shimmer \cite{ncvs.org}
\end{itemize} & 
\begin{itemize}[label={--},noitemsep,leftmargin=*,topsep=0pt,partopsep=0pt]
	\item jitterLocal\_sma\_amean
    \item shimmerLocal\_sma\_amean
\end{itemize}
\\
\hline
Glottalized & clicking noise heard during voicing & forceful adduction or abduction of the vocal folds during speech & 
\begin{itemize}[label={--},noitemsep,leftmargin=*,topsep=0pt,partopsep=0pt]
    \item Spectral Tilt \cite{hanson1999glottal}
    \item Increased Downward Spectral Slope \cite{gobl1988b}\cite{gobl1989preliminary}
    \item Decreased F1 Amplitude \cite{gobl1988b}\cite{gobl1989preliminary}
\end{itemize} & 
\begin{itemize}[label={--},noitemsep,leftmargin=*,topsep=0pt,partopsep=0pt]
	\item pcm\_fftMag\_mfcc\_sma[0]
    \item pcm\_fftMag\_mfcc\_sma[1]
    \item pcm\_fftMag\_mfcc\_sma[2]
    \item pcm\_fftMag\_mfcc\_sma[3]
	\item F1amplitudeLogRelF0
    \item pcm\_fftMag\_spectralSlope
\end{itemize}
\\
\\
\hline
Raspy or Hoarse & harsh, grating sound & combination of irregularity in vocal fold vibration and glottal noise generation &  
\begin{itemize}[label={--},noitemsep,leftmargin=*,topsep=0pt,partopsep=0pt]
    \item Pitch Amplitude (PA) - supposed to be lower for hoarse voices \cite{doi:10.1044/jshr.3302.298}
    \item Lower HNR \cite{doi:10.1044/jshr.3302.298} \cite{yumoto1982harmonics}
    \item Jitter and Shimmer \cite{pinto1990unification}\cite{kasuya1986acoustic}\cite{hollien1973method}\cite{koike1973application}
\end{itemize} & 
\begin{itemize}[label={--},noitemsep,leftmargin=*,topsep=0pt,partopsep=0pt]
	\item logHNR\_sma\_amean
	\item jitterLocal\_sma\_amean
    \item shimmerLocal\_sma\_amean
\end{itemize}
\\
\hline
Nasal or Honky & excessive nasality & excessive acoustic energy couples to the nasal tract & 
\begin{itemize}[label={--},noitemsep,leftmargin=*,topsep=0pt,partopsep=0pt]
    \item Increase in power level between the first and second formant. \cite{kataoka1996spectral}\cite{kataoka2001relationship}\cite{yoshida2000spectral}
    \item Decrease in power level in second and third formants region. \cite{kataoka1996spectral}\cite{kataoka2001relationship}\cite{yoshida2000spectral}
    \item Higher amplitudes for the bands centered at 630, 800 and 1000 Hz. \cite{lee2003acoustic}
    \item Significantly lower amplitude for the band centered at 2500 Hz. \cite{lee2003acoustic}
    \item One-third-octave analysis. \cite{kataoka2001relationship}\cite{kataoka1996spectral}
    \item Decrease in the amplitude of the first formant (F1) - Reduced F1 amplitude is associated with the increase in formant bandwidths and upward shifts in formant frequencies. \cite{lee2003acoustic}
    
\end{itemize} &
\begin{itemize}[label={--},noitemsep,leftmargin=*,topsep=0pt,partopsep=0pt]
    \item F1amplitudeLogRelF0
    \item F2amplitudeLogRelF0
   	\item F3amplitudeLogRelF0
\end{itemize}\\
\\
\hline
Jitter & pitch sounds rough & fundamental frequency varies from cycle to cycle &  Jitter/Jita & jitterLocal\_sma\_amean \\ 
\hline
Pressed & harsh, often loud (strident) quality & vocal processes of the arytenoid cartilages are squeezed together, constricting the glottis, and causing low airflow and medial compression of the vocal folds &
\begin{itemize}[label={--},noitemsep,leftmargin=*,topsep=0pt,partopsep=0pt]
    \item F0 not low \cite{keating2015acoustic}
    \item Glottal constriction (low H1-H2) \cite{keating2015acoustic}
\end{itemize} & 
\begin{itemize}[label={--},noitemsep,leftmargin=*,topsep=0pt,partopsep=0pt]
    \item 
\end{itemize}
\begin{itemize}[label={--},noitemsep,leftmargin=*,topsep=0pt,partopsep=0pt]
    \item F0final\_sma\_amean
    \item logRelF0-H1-H2\_sma3nz\_amean
\end{itemize}
\\
\hline
Pulsed or Strohbass or Vocal Fry & sounds similar to food cooking in a hot frying pan & sound gaps caused by intermittent energy packets below 70 Hz and formant energy dies out prior to re-excitation & 
\begin{itemize}[label={--},noitemsep,leftmargin=*,topsep=0pt,partopsep=0pt]
    \item Lower Pitch Amplitude (PA) \cite{doi:10.1044/jshr.3302.298}
    \item Lower HNR \cite{doi:10.1044/jshr.3302.298}\cite{keating2015acoustic}
    \item Low F0 \cite{keating2015acoustic}\cite{hollien1972vocal}
    \item F0 less than 70 Hz. \cite{titze1998principles}
    \ Glottal constriction (low H1-H2) \cite{keating2015acoustic}
\end{itemize} & 
\begin{itemize}[label={--},noitemsep,leftmargin=*,topsep=0pt,partopsep=0pt]
    \item logHNR\_sma\_amean
    \item F0final\_sma\_amean
    \item logRelF0-H1-H2\_sma3nz\_amean
\end{itemize}
\\
\hline
Resonant & brightened or 'ringing' sound that carries well & epilaryngeal resonance is enhanced, producing a strong spectral peak at 2500-3500 Hz; in effect, formants F3, F4 and F5 are clustered &
\begin{itemize}[label={--},noitemsep,leftmargin=*,topsep=0pt,partopsep=0pt]
    \item Not breathy \cite{titze2001acoustic} 
    \item Not pressed \cite{titze2001acoustic} 
    \item Low formant frequencies \cite{barrichelo2009resonant}
    \item Significant difference between F1 and H2 for the male \cite{barrichelo2009resonant}
    \item Significantly small difference between F1 and F0 for the female \cite{barrichelo2009resonant}
\end{itemize} & 
\begin{itemize}[label={--},noitemsep,leftmargin=*,topsep=0pt,partopsep=0pt]
    \item openSMILE correlates of breathy voice from ~\ref{tab:1} 
    \item openSMILE correlates of pressed voice from ~\ref{tab:1} 
    \item F1frequency\_sma3nz\_amean
    \item F2frequency\_sma3nz\_amean
    \item F3frequency\_sma3nz\_amean
\end{itemize}
\\
\hline
Rough & uneven, bumpy sound appearing to be unsteady short-term, but persisting over the long-term or presence of a low-frequency noise component & modes of vibration of the vocal folds are not synchronized & 
\begin{itemize}[label={--},noitemsep,leftmargin=*,topsep=0pt,partopsep=0pt]
    \item Pitch Amplitude (PA) \cite{doi:10.1044/jshr.3302.298}
    \item Lower HNR \cite{doi:10.1044/jshr.3302.298}
    \item Lower SFR (the flatness of magnitude spectrum (in dB) is the log of the ratio of the Geometric mean of the spectrum to Arithmetic mean of the spectrum \cite{doi:10.1044/jshr.3302.298}
    \item Mean HNR in b1 (400-2KHz) and b2 (2KHz-5KHz) bands \cite{de1995some}\cite{wolfe1987prediction}\cite{yumoto1982harmonics}
\end{itemize} &
\begin{itemize}[label={--},noitemsep,leftmargin=*,topsep=0pt,partopsep=0pt]
    \item logHNR\_sma\_amean
    \item audSpec\_Rfilt\_sma\_flatness
\end{itemize}
\\
\hline
Shimmer & crackly, buzzy & short-term (cycle-to-cycle) variation in a signal's amplitude & shimmer & shimmerLocal\_sma\_amean \\
\hline
Strained & effortfulness apparent in voice, hyperfunction of neck muscles, entire larynx may compress & excessive energy focused in laryngeal region & 
\begin{itemize}[label={--},noitemsep,leftmargin=*,topsep=0pt,partopsep=0pt]
    \item Cepstral peak prominence (CPP) \cite{fraile2014cepstral}\cite{lowell2012spectral}
    \item L/H ratio (spectral) and its standard deviation\cite{lowell2012spectral}
    \item Spectral Skewness/Tilt \cite{lowell2011spectral}
\end{itemize} & 
\begin{itemize}[label={--},noitemsep,leftmargin=*,topsep=0pt,partopsep=0pt]
    \item pcm\_fftMag\_spectralSkewness
    \item any energy based feature for computing L/H (e.g. pcm\_RMSenergy)
\end{itemize} \\
\hline
Tremerous & affected by trembling or tremors & modulation of 1-15 Hz in either amplitude or pitch due to a neurological or biomechanical cause &
\begin{itemize}[label={--},noitemsep,leftmargin=*,topsep=0pt,partopsep=0pt]
    \item Percent Jitter \cite{shao2010acoustic}
    \item Percent Shimmer \cite{shao2010acoustic}
\end{itemize} & 
\begin{itemize}[label={--},noitemsep,leftmargin=*,topsep=0pt,partopsep=0pt]
	\item jitterLocal\_sma\_amean
    \item shimmerLocal\_sma\_amean
\end{itemize}
\\
\hline
Twangy & sharp, bright sound & often attributed to excessive nasality, but probably also has an epilaryngeal basis &  F1 and F2 tend to be spread far apart for twangy vowels \cite{story2001relationship} & 
\begin{itemize}[label={--},noitemsep,leftmargin=*,topsep=0pt,partopsep=0pt]
    \item F1Frequency\_sma3nz\_amean 
    \item F2Frequency\_sma3nz\_amean
    \item F1bandwidth\_sma3nz\_amean
    \item F2bandwidth\_sma3nz\_amean
\end{itemize}\\
\hline
Ventricular or harsh & very rough (Louis Armstrong-type voice) & phonation using the false folds anterior rather than the vocal folds; unless intentional due to damage to the true folds, considered an abnormal muscle pattern dysphonia & 
\begin{itemize}[label={--},noitemsep,leftmargin=*,topsep=0pt,partopsep=0pt]
    \item F0 range twice that of the range for the non-harsh voice \cite{bowler1964fundamental}
    \item Distribution of F0 markedly skewed toward the lower end of the distribution \cite{bowler1964fundamental}
    \item Abrupt and extreme changes in the fundamental frequency (Jitter) \cite{bowler1964fundamental}
\end{itemize} & 
\begin{itemize}[label={--},noitemsep,leftmargin=*,topsep=0pt,partopsep=0pt]
    \item F0final\_sma\_range
    \item F0final\_sma\_peakDistStddev
    \item jitterLocal\_sma\_amean
\end{itemize}\\
\\
\hline
Wobble & wavering or irregular variation in sound & amplitude and/or frequency modulations in the 1-3 Hz range &  FM and/or AM $\leq$ 3 Hz &
\begin{itemize}[label={--},noitemsep,leftmargin=*,topsep=0pt,partopsep=0pt]
    \item jitterLocal\_sma\_amean
    \item shimmerLocal\_sma\_amean
\end{itemize}\\
\\
\hline
Yawny & quality is akin to sounds made during a yawn & larynx is lowered and pharynx is widened, as people do when yawning - hence the name & formants F1 and F2 in the yawn vowel quality tend to be more closely spaced than in the normal vowels \cite{story2001relationship} &
\begin{itemize}[label={--},noitemsep,leftmargin=*,topsep=0pt,partopsep=0pt]
    \item F1Frequency\_sma3nz\_amean 
    \item F2Frequency\_sma3nz\_amean
    \item F1bandwidth\_sma3nz\_amean
    \item F2bandwidth\_sma3nz\_amean
\end{itemize}\\
\bottomrule
\end{longtable}

\section{Inference of Speaker Traits and States from Acoustic Cues}
\label{sec:paralinguistictable}
Table~\ref{tab:2} defines the mapping from acoustic correlates or voice qualifiers to paralinguistic inferences researchers have made over the years across different literature.
\begin{longtable}{
p{1cm}
p{5.5cm}
p{7cm}
p{1cm}
}
\caption{Paralinguistic Inferences from Voice}\label{tab:2}\\
 \toprule  
\multicolumn{1}{c}{Meta-Inference}&
\multicolumn{1}{c}{Acoustic or Perceptual Correlate(s)}&
\multicolumn{1}{c}{openSMILE Correlate(s)}&
\multicolumn{1}{c}{Source}\\
\midrule
\endfirsthead
\caption[]{Paralinguistic Inferences from Voice}\\
 \toprule  
\multicolumn{1}{c}{Meta-Inference}&
\multicolumn{1}{c}{Acoustic or Perceptual Correlate(s)}&
\multicolumn{1}{c}{openSMILE Correlate(s)}&
\multicolumn{1}{c}{Source}\\
\midrule
\endhead

Confidence & 
\begin{itemize}[label={--},noitemsep,leftmargin=*,topsep=0pt,partopsep=0pt]
    \item Greater spectral energy 
    \item Greater Pitch
    \item Faster rate of speech
    \item Less number of pauses
    \item Less duration of pauses in comparison to doubtful voice
\end{itemize} & 
\begin{itemize}[label={--},noitemsep,leftmargin=*,topsep=0pt,partopsep=0pt]
    \item pcm\_RMSenergy 
    \item pcm\_zcr
    \item F0
\end{itemize}
& \cite{scherer1973voice} \\
\hline
Enthusiastic, Forceful, Active, Competent & Confidence &
\begin{itemize}[label={--},noitemsep,leftmargin=*,topsep=0pt,partopsep=0pt]
    \item pcm\_RMSenergy 
    \item pcm\_zcr
    \item F0
\end{itemize}
& \cite{scherer1973voice} \\
\hline
Doubtful & 
\begin{itemize}[label={--},noitemsep,leftmargin=*,topsep=0pt,partopsep=0pt]
    \item Less spectral energy 
    \item Lower Pitch
    \item Slower rate of speech
    \item More number of pauses
    \item Longer pauses 
\end{itemize}
& 
\begin{itemize}[label={--},noitemsep,leftmargin=*,topsep=0pt,partopsep=0pt]
    \item pcm\_RMSenergy 
    \item pcm\_zcr
    \item F0
\end{itemize}
& \cite{scherer1973voice} \\
\hline
Warmth (Male) & 
\begin{itemize}[label={--},noitemsep,leftmargin=*,topsep=0pt,partopsep=0pt]
    \item Higher F0 range 
    \item Higher Mean Spectral Slope (0-500Hz)
    \item Lower Standard Deviation (SD) of F1 and F2
\end{itemize}
& 
\begin{itemize}[label={--},noitemsep,leftmargin=*,topsep=0pt,partopsep=0pt]
    \item F0final\_sma\_range
    \item pcm\_fftMag\_spectralSlope
    \item slopeV0\-500\_sma3nz\_amean
    \item slopeUV0\-500\_sma3nz\_amean
    \item F1frequency\_sma3nz\_stddevNorm
    \item F2frequency\_sma3nz\_stddevNorm
\end{itemize}
& \cite{gallardo2017perceived}\cite{eyben2016geneva} \\
\hline
Attractiveness (Male) & 
\begin{itemize}[label={--},noitemsep,leftmargin=*,topsep=0pt,partopsep=0pt]
    \item Higher F0 range 
    \item Higher Standard Deviation (SD) of Hammarberg Index
    \item Lower mean length of unvoiced segments
    \item Higher SD of F0
    \item Lower median of F0
\end{itemize}
&
\begin{itemize}[label={--},noitemsep,leftmargin=*,topsep=0pt,partopsep=0pt]
    \item F0final\_sma\_range
    \item hammarbergIndexV\_sma3nz\_stddevNorm
    \item hammarbergIndexUV\_sma3nz\_stddevNorm
    \item voicingFinalUnclipped
    \item F0final\_sma\_stddev
    \item F0semitoneFrom27.5Hz\_sma3nz\_stddevNorm
    \item F0semitoneFrom27.5Hz\_sma3nz\_percentile50.0
\end{itemize}
& \cite{gallardo2017perceived}\cite{eyben2016geneva} \\
\hline
Confidence (Male) & 
\begin{itemize}[label={--},noitemsep,leftmargin=*,topsep=0pt,partopsep=0pt]
    \item Lower median F0 
    \item Higher F0 range
\end{itemize}
&
\begin{itemize}[label={--},noitemsep,leftmargin=*,topsep=0pt,partopsep=0pt]
    \item F0final\_sma\_range
    \item F0semitoneFrom27.5Hz\_sma3nz\_percentile50.0
\end{itemize}
& \cite{gallardo2017perceived}\cite{eyben2016geneva} \\
\hline
Compliance (Male) & 
Lower SD length of voiced segments
&
voicingFinalUnclipped
& \cite{gallardo2017perceived}\cite{eyben2016geneva} \\
\hline
Maturity (Male) & 
\begin{itemize}[label={--},noitemsep,leftmargin=*,topsep=0pt,partopsep=0pt]
    \item Lower median F0 
    \item Higher SD of F3
    \item Higher SD of F3 bandwidth
\end{itemize}
&
\begin{itemize}[label={--},noitemsep,leftmargin=*,topsep=0pt,partopsep=0pt]
    \item F0semitoneFrom27.5Hz
    \item F3frequency\_sma3nz\_stddevNorm
    \item F3bandwidth\_sma3nz\_stddevNorm
\end{itemize}
& \cite{gallardo2017perceived}\cite{eyben2016geneva} \\
\hline
Warmth (Female) & 
\begin{itemize}[label={--},noitemsep,leftmargin=*,topsep=0pt,partopsep=0pt]
    \item Higher F0 range
    \item Higher F1
    \item Higher SD of F2 bandwidth
    \item Higher SD of spectral flux
    \item Lower SD of F1
    \item Higher median F0
\end{itemize}
&
\begin{itemize}[label={--},noitemsep,leftmargin=*,topsep=0pt,partopsep=0pt]
    \item F0final\_sma\_range
    \item F1frequency\_sma3nz\_amean
    \item F2bandwidth\_sma3nz\_stddevNorm
    \item pcm\_fftMag\_spectralFlux
    \item F1frequency\_sma3nz\_stddevNorm
    \item F0semitoneFrom27.5Hz
\end{itemize}
& \cite{gallardo2017perceived}\cite{eyben2016geneva} \\
\hline
Attractiveness (Female) & 
\begin{itemize}[label={--},noitemsep,leftmargin=*,topsep=0pt,partopsep=0pt]	
	\item Higher F0 range
    \item Higher F1
    \item Higher SD of F2 bandwidth
    \item Lower SD of F1
    \item Lower mean spectral slope (0-500 Hz)
\end{itemize}
&
\begin{itemize}[label={--},noitemsep,leftmargin=*,topsep=0pt,partopsep=0pt]
    \item F0final\_sma\_range
    \item F1frequency\_sma3nz\_amean
    \item F2bandwidth\_sma3nz\_stddevNorm
    \item F1frequency\_sma3nz\_stddevNorm
    \item slopeV0\-500\_sma3nz\_amean
    \item slopeUV0\-500\_sma3nz\_amean
\end{itemize}
& \cite{gallardo2017perceived}\cite{eyben2016geneva} \\
\hline
Confidence (Female) & 
\begin{itemize}[label={--},noitemsep,leftmargin=*,topsep=0pt,partopsep=0pt]
	\item Higher F0 range
    \item Higher SD of falling slope of loudness
\end{itemize}
&
\begin{itemize}[label={--},noitemsep,leftmargin=*,topsep=0pt,partopsep=0pt]
    \item F0final\_sma\_range
    \item loudness\_sma3\_stddevFallingSlope
\end{itemize}
& \cite{gallardo2017perceived}\cite{eyben2016geneva} \\
\hline
Compliance (Female) & 
\begin{itemize}[label={--},noitemsep,leftmargin=*,topsep=0pt,partopsep=0pt]
    \item Lower SD of F1
    \item Higher F1
    \item Lower loudness range
\end{itemize}
&
\begin{itemize}[label={--},noitemsep,leftmargin=*,topsep=0pt,partopsep=0pt]
    \item F1frequency\_sma3nz\_stddevNorm
    \item F1frequency\_sma3nz\_amean
    \item loudness\_sma3\_amean
    \item loudness\_sma3\_stddevNorm
\end{itemize}
& \cite{gallardo2017perceived}\cite{eyben2016geneva} \\
\hline
Maturity (Female) & 
\begin{itemize}[label={--},noitemsep,leftmargin=*,topsep=0pt,partopsep=0pt]
    \item Lower median F0
    \item Higher mean mfcc4
    \item Lower F1
    \item Higher mean mfcc2
\end{itemize}
&
\begin{itemize}[label={--},noitemsep,leftmargin=*,topsep=0pt,partopsep=0pt]
    \item F0semitoneFrom27.5Hz or F0final
    \item mfcc4\_sma3\_amean
    \item F1frequency\_sma3nz\_amean
    \item mfcc2\_sma3\_amean
\end{itemize}
& \cite{gallardo2017perceived}\cite{eyben2016geneva} \\
\hline
Dysphonia & 
Strained Voice Quality
&
Acoustic correlates of strained voice quality from ~\ref{tab:1}
& \cite{eadie2005classification}\cite{heman2002relationship}\cite{heman2003cepstral} \\
\hline
Vocal Tremor & 
Tremerous Voice Quality
&
Acoustic correlates of tremerous voice quality from ~\ref{tab:1}
& \cite{garellek2012timing} \\
\hline
Authoritative Personality & 
Confidence, Creakiness
&
Acoustic correlates of confidence, and creaky voice quality from ~\ref{tab:2} and ~\ref{tab:1} respectively
& \cite{hildebrand2016creaky}\cite{steinmetz_2011}\cite{yuasa2010creaky} \\
\hline
Likability (male) & 
\begin{itemize}[label={--},noitemsep,leftmargin=*,topsep=0pt,partopsep=0pt]
    \item Lower F0
    \item s\_cog (center of gravity for spectrum averaged for each stimulus)
\end{itemize}
&
\begin{itemize}[label={--},noitemsep,leftmargin=*,topsep=0pt,partopsep=0pt]
    \item F0semitoneFrom27.5Hz or F0final
    \item audSpec\_Rfilt\_sma\_centroid
\end{itemize}
& \cite{weiss2010voice} \\
\hline
Likability (female) & 
higher s\_sd (standard deviation for spectrum averaged for each stimulus)
&
spectralFlux\_sma3\_amean
& \cite{weiss2010voice} \\
\hline
Dominance (male) & 
Lower F0 variation
&
F0semitoneFrom27.5Hz or F0final
& \cite{hodges2010different} \\
\hline
Happiness, Joy, Confidence, Anger, Fear & 
High Pitch levels
&
F0\_mean
& \cite{hodges2010different} \\
\hline
\bottomrule
\end{longtable}

\bibliography{ref}
\bibliographystyle{unsrt}

\end{document}